\documentclass{aa}
\usepackage{graphicx}
\usepackage{natbib}
\bibpunct{(}{)}{;}{a}{}{,} 
\usepackage{txfonts}
\begin{document}
\title{The thick disk rotation-metallicity correlation as a fossil  of
  an ``inverse chemical gradient'' in the early Galaxy} 

   \author{A. Curir\inst{1},
           M. G. Lattanzi\inst{1},
         A. Spagna\inst{1},
   F. Matteucci\inst{2},
   G. Murante\inst{1},
          P. Re Fiorentin\inst{1},
          E. Spitoni\inst{3}
          }
   \institute{INAF--Osservatorio Astronomico di Torino,
             via Osservatorio 20, 10025 Pino Torinese, Italy\\
             \email{curir@oato.inaf.it}
        \and
              Department of Mathematics, Physics and Natural Sciences, University of Trieste       
             \and
  Department of Mathematics, University of Evora, Portugal}

   \authorrunning{A. Curir et al.}

  \abstract
{The thick disk  rotation--metallicity correlation, $\partial V_\phi/\partial$[Fe/H]$=40\div 50$ km~s$^{-1}$dex$^{-1}$ represents an important 
signature of the formation processes of the galactic disk.}
{ We use nondissipative numerical simulations to follow the evolution of a Milky Way (MW)-like 
disk to verify if  secular dynamical processes can account for this correlation 
in the old thick disk stellar population.} 
{We followed the evolution of an ancient disk population represented by 10 million particles 
whose chemical abundances were assigned by assuming a  cosmologically plausible radial metallicity gradient with
lower metallicity in the inner regions, as expected for the 10-Gyr-old MW. 
The two cases of a disk with and without a bar  were  simulated  to compare the evolution 
of their kinematics and radial chemical properties.}
{Migration processes act in both cases and  appear to be enhanced in the presence of a central bar. 
Essentially, inner disk stars move towards the outer regions and populate layers located at higher $|z|$.  
In the case of an evolved barred disk, a  rotation--metallicity correlation appears, 
which  well  resembles  the behaviour observed in our Galaxy at a
galactocentric distance between  8 kpc and 10 kpc. In particular,
we measure a correlation of 
$\partial V_\phi/\partial$[Fe/H]$\simeq  60 $ km~s$^{-1}$dex$^{-1}$ 
for particles  at 1.5 kpc $< |z| < 2.0$ kpc that persists up to 6 Gyr.}
{ 
{ Our pure $N$-body models can account for the $V_\phi$ vs. [Fe/H] correlation observed in the 
 thick disk of our Galaxy, suggesting that  processes internal to the disk such as heating and radial migration  play a role 
in the formation of this old stellar component. In this scenario, the positive rotation-metallicity
correlation of the  old thick disk population would represent  the
relic signature of an ancient {{\it inverse}} chemical (radial) gradient in the
inner Galaxy, which resulted from  accretion of primordial gas.}}

   \maketitle
%

\section{Introduction}
Two distinct disk populations are observed  in the solar neighbourhood
in terms of kinematics, average metallicities, and stellar ages.
They belong to a {\it thin disk}, younger than 7-8 Gyr, and to a {\it thick disk}, older
than 8-9 Gyr 
\citep[e.g.][and references therein] {haywood2008}.\\
We now  have  several different  scenarios for the origin of the thick disk:
kinematical heating of a pre-existing old disk via minor mergers
 \citep{quinn1993, robin1996, villalobos2008, Qu2011},  accretion of satellites that have deposited their stellar debris in planar configuration \citep{abadi2003},
  gas accretion at high redshift and stars formed in situ \citep{brook2005}, wet merger \citep{martel2011}, clumpy disks \citep{bournaud2009}, and radial migration  \citep{schonrich2009a,schonrich2009b,loebman2011}. The interested reader can consult the work by \cite{lee2011} for a detailed overview of the  possible origins of the thick disk.\\
Chemical properties of stars provide important clues to disentangle the puzzle of the Galaxy formation. By measuring
 abundance ratios of stars in different
parts of the Galaxy, one can infer how fast
 metal enrichment proceeded and the timescale over
which the different regions were formed.  The distribution of the chemical elements in our
Galaxy can be considered as a "fossil record" of its evolutionary
history. Therefore crucial information on the mechanism that dominated  the
formation of the thick disk  is encoded in the chemical properties of its
stars. By comparing the chemical properties of bulge, thick and thin disk
stars, and by correlating them with other kinematical data, one can identify
the processes  that
 play a dominant role in the formation of the different Galactic components.\\
On this matter, an intriguing correlation between thick disk rotation and
metallicity was found by \cite{spagna2010}. Subsequently, this correlation was confirmed by \cite{lee2011}
using similar data  and, very recently, by \cite{kordopatis2011} using fully independent spectroscopic data.
The Spagna et al. correlation  is radically different from what is observed for
the thin disk stars \citep{haywood2008},  and from  predictions made by the  chemodynamical models
of \cite{schonrich2009a} and of \cite{loebman2011}  for young   stars.\\
In the context of  stellar migration models, the negative  rotation--metallicity correlation shown by the thin disk stars in the solar neighbourhood derives 
from the ''immigration'' of slower rotating stars that come from the inner disk regions and are characterised by stellar populations with
higher chemical abundances. \\
Conversely, the positive correlation shown by the $V_\phi$ vs. [Fe/H] distribution of thick disk stars observed {\it in situ} above 1 kpc from the plane could be explained with a population of much older stars that come from the inner 
regions of the Galaxy and have formed during the early disk star formation phases 
from molecular clouds with relatively low chemical abundance and higher $\alpha$-enrichment.
Note  that  relationships between  thick disks and  bulges 
in external galaxies have been known for quite some time \citep{Freeman 1987}. \\
 Although this substantially positive $V_\phi$ vs. [Fe/H] correlation  has not yet received much theoretical attention, a weak positive correlation is found in the simulations carried out by  \citet{loebman2011},
while the statistical model developed by \cite{schonrich2009a} indicates only a mild trend 
 (10 km s$^{-1}$ dex$^{-1}$) at z =0 kpc, which decreases with height and disappears for $|z| > 1$ kpc (Sch\"{o}nrich 2009, private communication).\\

 In this paper we  show  that if we assume the early radial chemical gradient
derived from  the inside-out formation and chemical evolution model of the
Galactic disk as originally suggested by \cite{matt89} and  \cite{chiap01}, a positive correlation between rotation
velocity and metallicity can be established as a result  of   radial migration and heating processes of stars from the inner region of the disk.  These   processes are investigated by evolving a stellar disk with  $N$- body simulations and thus following how  the  chemical properties assigned to the initial configuration are redistributed.\\
 Our model cannot be
viewed as a complete galaxy evolution model, because we neglected the
\emph{gradual} formation and growth of the stellar disk. However,
our approach allows us to  
 analyse the dynamical evolution of the non dissipative component, and, therefore,  to separate the   chemical evolution from 
 dynamics, i. e. leaving any redistribution of chemical gradients to stellar
 motion.

\section{Observed rotation -- metallicity correlation}
\label{sect:correlation}
 The observed rotation--metallicity correlation  is based on a new kinematic catalogue derived by assembling the astrometric 
parameters extracted from the database used for the construction of the Second Guide Star Catalog \citep[GSC-II; ][]{lasker2008} 
with spectro-photometric data from the Seventh Data Release of the Sloan Digital Sky Survey \citep[SDSS DR7; see e.g.\ ][]{abazajian2009, yanny2009}.  
The SDSS--GSC-II catalogue contains positions, proper motions, classification, and $ugriz$ photometry for 77 million sources down to $r\approx 20$,  over 9000 square-degrees. 
\cite{spagna2010} analysed 27\,000 FGK (sub)dwarfs in the solar neighbourhood and found evidence of a  rotation--metallicity correlation, $\partial V_\phi / \partial$[Fe/H]$\approx 40\div 50$ km~s$^{-1}$ dex$^{-1}$, amongst {\it in situ} thick disk stars, located at 1 kpc $<|z|<3$ kpc and selected with abundance $-1.0<$[Fe/H]$<-0.5$. The thin disk contamination is expected to be negligible in this sample; however, this is not true for the halo stars, consequently, the thick disk and halo populations had to be deconvolved by assuming the  superposition of two Gaussian velocity distributions. 

This result was also confirmed over a wider height interval, $0.1$ kpc $<|z|< 3$ kpc, by \cite{lee2011}, who analysed the kinematics of $\sim$ 17\,500 G-type dwarfs from which they selected a subsample of {\it bona fide} thick disk stars by means of [Fe/H] and [$\alpha/$Fe] provided by the spectroscopic SDSS sample.  The fact that other studies based on the photometric SDSS sample (\cite{ivezic2008}, \cite{bond2010}, and \cite{loebman2011}) did not reveal any significant rotation-metallicity correlation seems to depend on the larger errors affecting the chemical abundances derived from the $ugriz$ passbands (\cite{arnadottir2010}, \cite{lee2011}). 
 For these reasons, it is of particular importance to cite \cite{kordopatis2011}, who recently performed a spectroscopic survey of 700 thick disk stars using fully independent VLT/FLAMES observations and 
 found the same correlation  within the interval 1 kpc $<|z|<$ 4 kpc from the plane, where the thick disk population dominates.
Finally, a positive correlation was present in the sample discussed in  \cite{gratton2003}, who analysed 150 subdwarfs and early subgiants in the solar neighbourhood with homogeneous  spectroscopic abundances and accurate kinematics derived from the Hipparcos catalogue.\\
   The discussion above clearly shows that part of the burden is still on the observational side and therefore on the decisive contribution 
that Gaia's  parallaxes will bring to this topic, along the lines discussed in \cite{freeman2002} and in \cite{refiorentin2012} for the MW halo.
  
  
\section{Chemical distribution}
\label{sect:metallicity}

 We computed the expected gradient in [Fe/H] along the disk of the MW at various cosmic epochs. To do that we ran a detailed chemical evolution 
model as described in \cite{spit01}, which is an upgraded version of the original two-infall model of Chiappini et al.(1997;2001), where details can be found. This model follows the evolution of 31 chemical species including H, He, D, C,N,O, $\alpha$-elements, Fe and Fe-peak elements. Stellar nucleosynthesis is taken into account in detail and includes the contributions from low - and intermediate - mass stars (0.8-8$M{\odot}$), massive stars ($M>8M_{\odot}$) that die as supernovae (SNe) of Type II  and Type Ib/c, and Type Ia SNe (white dwarfs in binary systems). One of the main assumptions of the model is that the thin disk of the Galaxy forms inside-out, namely the inner disk regions are assembled first by gas accretion of extragalactic origin (primordial chemical composition) and  then the outer disk regions form on timescales that vary from 2 Gyr in the inner disk (2 kpc) to 10 Gyr and more at larger galactocentric distances ($<$14 kpc).
This ensures the formation of a gradient especially if coupled with a threshold in the gas density regulating star formation. Radial flows were not considered \footnote{Including radial flows is necessary to generate abundance gradients only if no inside-out and no threshold are assumed (Spitoni \& Matteucci, 2011)}.

 The model adopted in this paper is Model S2IT of \cite{spit01} ( their Table 1):  it is a two-infall model similar to that presented in Chiappini et al. (2001) and  \cite{cesc07}. This particular model was chosen because it accurately reproduces the most recent estimate of the present time abundance gradients by \cite{luck2011} along the disk, who found $ \Delta$[Fe/H]$/ \Delta R=-0.062 \pm 0.002$ dex kpc$^{-1}$. Our predicted present time gradient for Fe is $\Delta$[Fe/H]$/ \Delta R=-0.052$ dex kpc$^{-1}$ in a galactocentric range 4-14 kpc. Given the uncertainties still existing in deriving chemical abundances, and in particular  in the adoption of different model atmospheres, and considering the small difference in the galactocentric distance range adopted by us compared to \cite{luck2011},  we consider our model predictions to be in reasonable agreement with the observations.

 \cite{spit01}  considered the gradient at 2 Gyr since the beginning of star formation and the one at 8 Gyr. While the gradient at 8 Gyr shows a clear increase of [Fe/H] with decreasing galactocentric distance, the gradient at 2 Gyr  shows an increase from the outer regions towards to 10 kpc, where it reaches a peak, and then a decrease from R$<$ 10 kpc, as shown in Figure 1. This apparently strange behaviour was already present in the model of Chiappini et al. (2001) and  can be explained on the basis of the inside-out disk formation, in particular, by the fact that at early epochs the efficiency of chemical enrichment in the inner regions of the disk is low  due to the very large amount of infalling primordial gas. Then, at later epochs, while the star formation rate is still much higher than in the outer disk regions, the infall of primordial gas is much stronger in the outer disk. This produces the early flat gradient and the steepening of the gradient in time.
 We adopted a metallicity dispersion, $\sigma_{\rm [Fe/H]}=0.1$ dex,
corresponding to 
the typical precision of the spectroscopic metallicity  measurements for SDSS, as estimated by \cite{allende2008}.
 
\begin{figure}[]
\resizebox{\hsize}{!}{\includegraphics{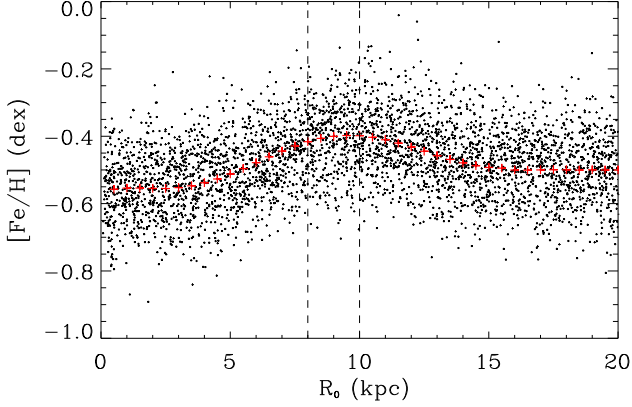}}
\caption{\footnotesize  Initial radial chemical distribution.  The crosses show the model of Spitoni \& Matteucci (2011), while the dots, which represent the chemical distribution of the N-body particles at $t=0$ Gyr, include an intrinsic dispersion $\sigma_{\rm [Fe/H]}=0.1$ dex.}  
\label{fig2}
\end{figure}

\section{Simulations}
\label{sec:sim}
We used a  dark matter (DM) halo containing a rotating stellar
exponential disk. The DM halo has a Navarro, Frenk and White  (NFW) \citep{NFW97} radial density profile
and structural parameters as indicated in Table 1. These parameters are appropriate for an MW-like DM
halo at redshift $z=0$.

\begin{table}
\caption[Table 1]{Halos properties (Navarro, Frenk and White profile)}
\begin{tabular}{*{8}{c}}
    \hline  DM  & $M_{vir}$ & $R_{vir}$ & $C_{vir}$ & $R_{max}$ & N & ${\epsilon}$ & $M_{DM}$\\
    \hline  {\it Halo}  &  $10^{12}$  & $258$ & $7.40$ & $336$ & $10^7$ & $0.11$  &
    $1.07\times 10^5$ \\
\hline
\end{tabular}
\begin{list}{}{}
\item[Notes.] $M_{vir}$: Halo's virial mass in $M_{\odot}$; $R_{vir}$: virial radius
  in kpc. $C_{vir}$: NFW concentration parameter. $R_{max}$: maximum
  radius. N: total number of Halo particles. ${\epsilon}$: softening
  length in kpc. $M_{DM}$: mass of DM particle in $M_{\odot}$.
\end{list}
\caption[Table 2]{Properties of the  disk}
  \begin{tabular}{*{8}{c}}
  \hline Stars  & $M_{*}$ & $M_{star}$ & $h_d$& $z_d$ & N & ${\epsilon}$ &$Q$\\
  \hline {\it Disk} & $5.6{\cdot}10^{10}$ & $7.47{\cdot}10^{3}$ & $3.5$ & $0.7$& $7.5
  {\cdot}10^{6}$ & $0.044$ & $2$\\
\hline
  \end{tabular}
\begin{list}{}{}
\item[Notes.] $M_{*}$:
  disk mass in $M_{\odot}$.  $M_{star}$: mass of the disk particle in
  $M_{\odot}$.  $h_d$: disk scale length in kpc. $z_d$: initial disk thickness. N: number of particles. ${\epsilon}$: softening length in kpc. $Q$: Toomre parameter.
\end{list}
\label{tab:disk}
\caption[Table 3]{Properties of the bulge (Hernquist profile)}
 \begin{tabular}{*{6}{c}}
  \hline Stars  & $M_{*}$ & N &   ${\epsilon}$ & $a$& $M_{b}$\\  
 \hline {\it Bulge} & $1.85{\cdot}10^{10}$ & $2.5{\cdot}10^{6}$ &  $0.044$ & $1.12$ & $7.39{\cdot}10^{3}$\\ 
\hline 
\end{tabular}
\begin{list}{}{}
\item[Notes.] $M_{*}$:mass in
  $M_{\odot}$. N: number of particles.  ${\epsilon}$: softening length in
  kpc. $a$: Hernquist scale radius in kpc. $M_{b}$: mass of the particle in $M_{\odot}$.
\end{list}
\end{table}
We used MW parameters at z=0 to initialise our model:
since we want to compare gradients observed  at present time with those generated by
dynamical evolution, we need to have the correct physical parameters of
today. Of course, this  introduces a discrepancy, because several Gyr ago the
Galaxy was different. However, we believe that  our  model, even if it is a simple nondissipative one,  is adequate to separate
 effects of dynamics from  chemical  evolution.
 The  DM particles of the nonrotating halo have velocities given by the local equilibrium
approximation \citep{hernquist1993}.\\
A stellar disk, whose mass is set to the
 value estimated for the MW, is embedded into our halo; the other features
of the simulated disk are described in Table 2. The position of  each
disk particle is obtained by  using the rejection method in
\cite{Press1986}; the disk is in gravitational equilibrium with the DM halo.
We also simulated  a  comparison case consisting of the same disk but endowed
with a bulge with  a
Hernquist radial density profile  and mass of $1.8 \times 10^{10} M_{\odot}$
(see Table 3).
The bulgeless disk decays in a barred configuration, while the disk with a bulge is
stabilised against  bar formation.
 We ran our two  simulations using the public parallel treecode GADGET2 \citep{sprin2005} on the cluster matrix at the CASPUR (Consorzio Interuniversitario per le Applicazioni del Supercalcolo) consortium, Rome.
Both  systems were left to evolve for 10 Gyr.  The snapshots of the particle configurations were taken with a timestep of $0.25 \times 10^8$ years. To avoid the effect of  relaxation of our initial conditions, we let the dynamical system to evolve for 1 Gyr before  assigning our initial metallicities\footnote{From this point on any reference to time evolution will be given with regard to the time of injection of the chemical distribution in the simulation.  Therefore $t=0$ corresponds to  1 Gyr from the beginning of the dynamical simulation.}, and we analysed the redistribution of the chemical properties  over the following 9 Gyr.
 We assigned  the   radial chemical distribution function defined in Sect. \ref{sect:metallicity}, which is shown in Figure \ref{fig2} to what we considered our initial configuration.  
Each particle in the  configuration is tagged with  a [Fe/H] label
according to this initial radial function.

\section{Results} 
\subsection{Evolution of  $V_\phi$ vs. [Fe/H]}
\begin{figure}
{\resizebox{\hsize}{!}{\includegraphics{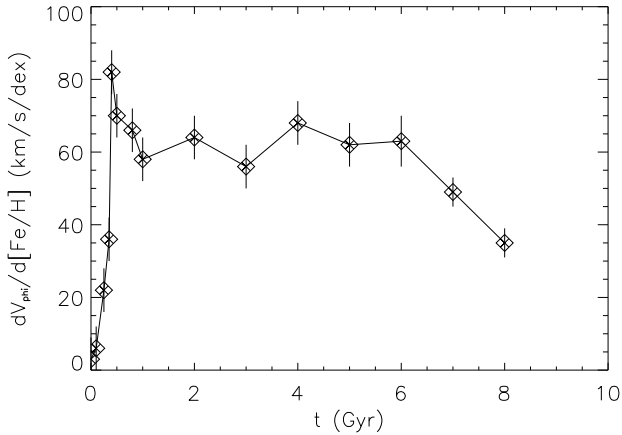}}}
\caption{Time evolution of the coefficient  $\partial V_\phi / \partial$[Fe/H]}
\label{ctempo}
\end{figure}

\begin{figure*}
\hspace{2.5truecm} \vbox{ 
    \includegraphics[scale=0.9]{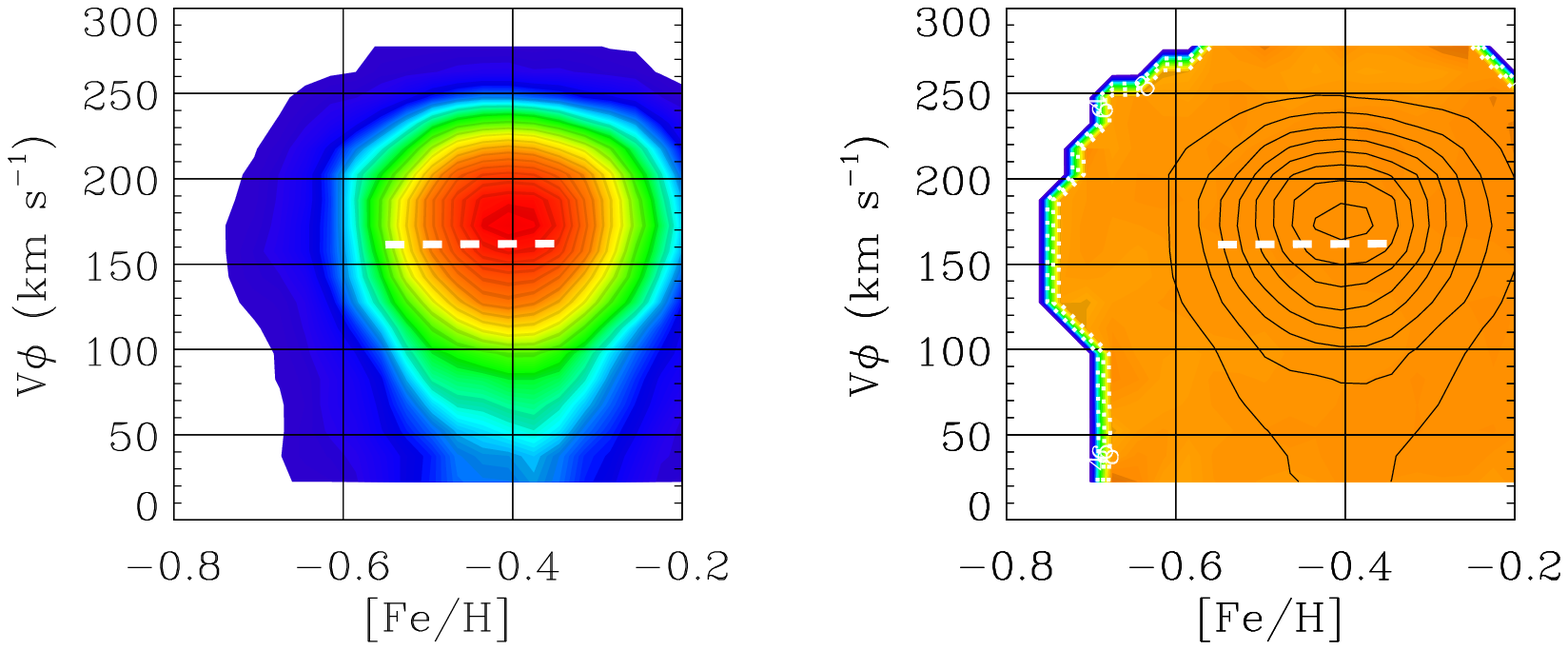}
    \includegraphics[scale=0.9]{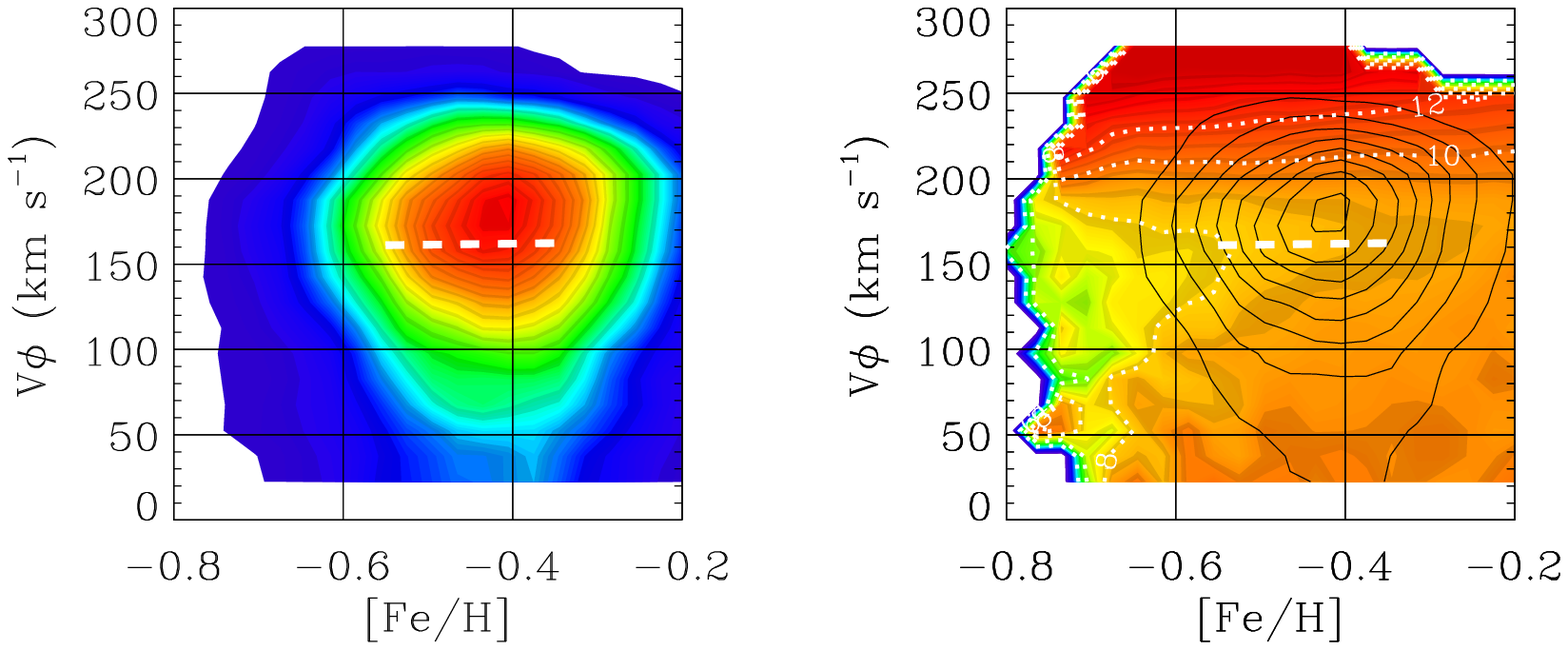}
    \includegraphics[scale=0.9]{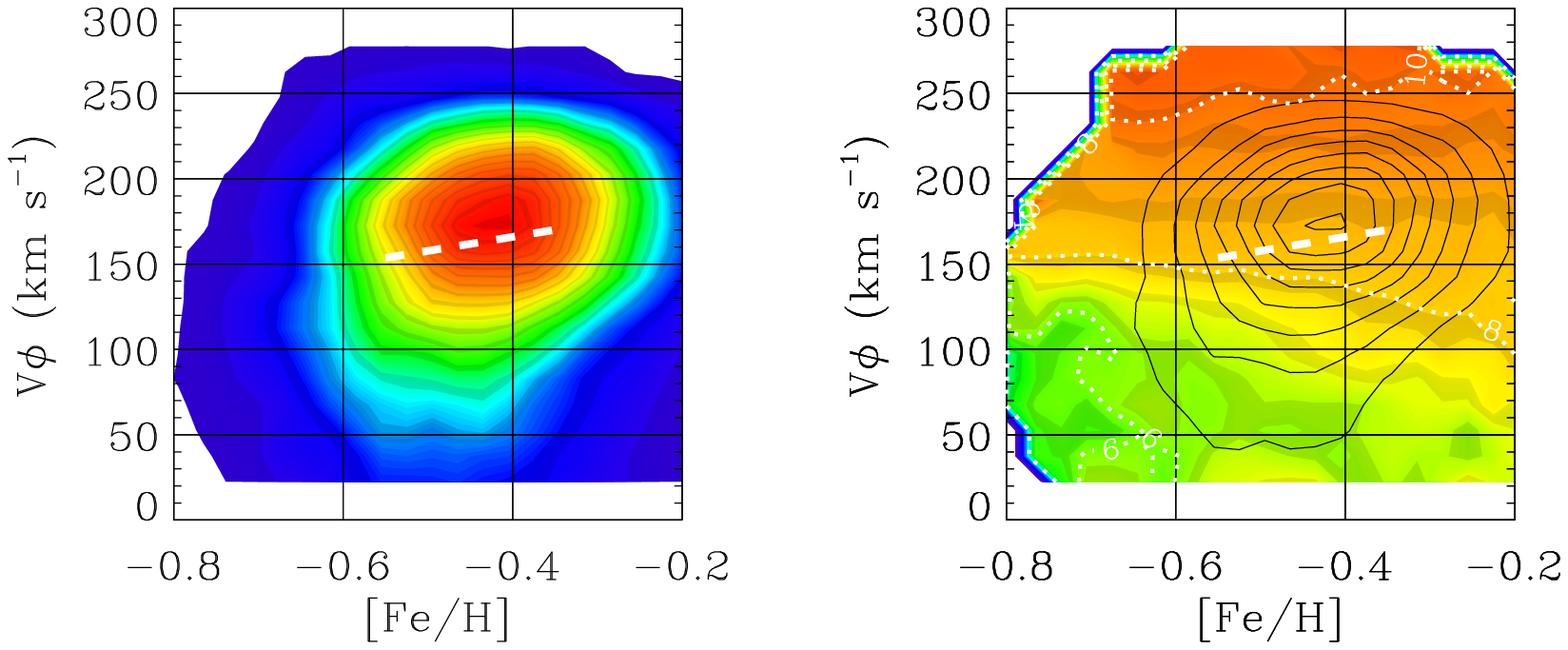} } 
\caption{\footnotesize
Barred-disk model. Distribution of the azimuthal velocity and metallicity after an evolution of 0, 0.1, 0.4  Gyr.
{\it Left panels}: colour contour plot showing the rotation-metallicity distribution, $V_\phi$ vs. [Fe/H], 
of the particles in the solar annulus, 8 kpc $<R<$ 10 kpc and 1.5 kpc $<|z|<$ 2.0 kpc. 
The dashed white line indicates the linear fit in the central region of the distribution, 
including all  particles with $-0.55< $ [Fe/H] $<-0.35$ and $V_\phi  > 50$ km~s$^{-1}$.
{\it Right panels}: the colour contour plot represents the distribution of the mean  original radius, 
$\langle {R_0} \rangle$, as a function of ([Fe/H],$V_\phi$) of the same particles shown in the left plot. 
Here, the dotted white lines mark the iso-contours at $\langle {R_0} \rangle$=6 kpc, 8 kpc, 10 kpc, and 12 kpc.
The black contours indicate the iso-density levels of the ([Fe/H],$V_\phi$) 
distribution shown in the left panel.}
\label{fig:correlation_n}
\end{figure*}

\begin{figure*}
\hspace{2.5truecm} \vbox{ 
    \includegraphics[scale=0.9]{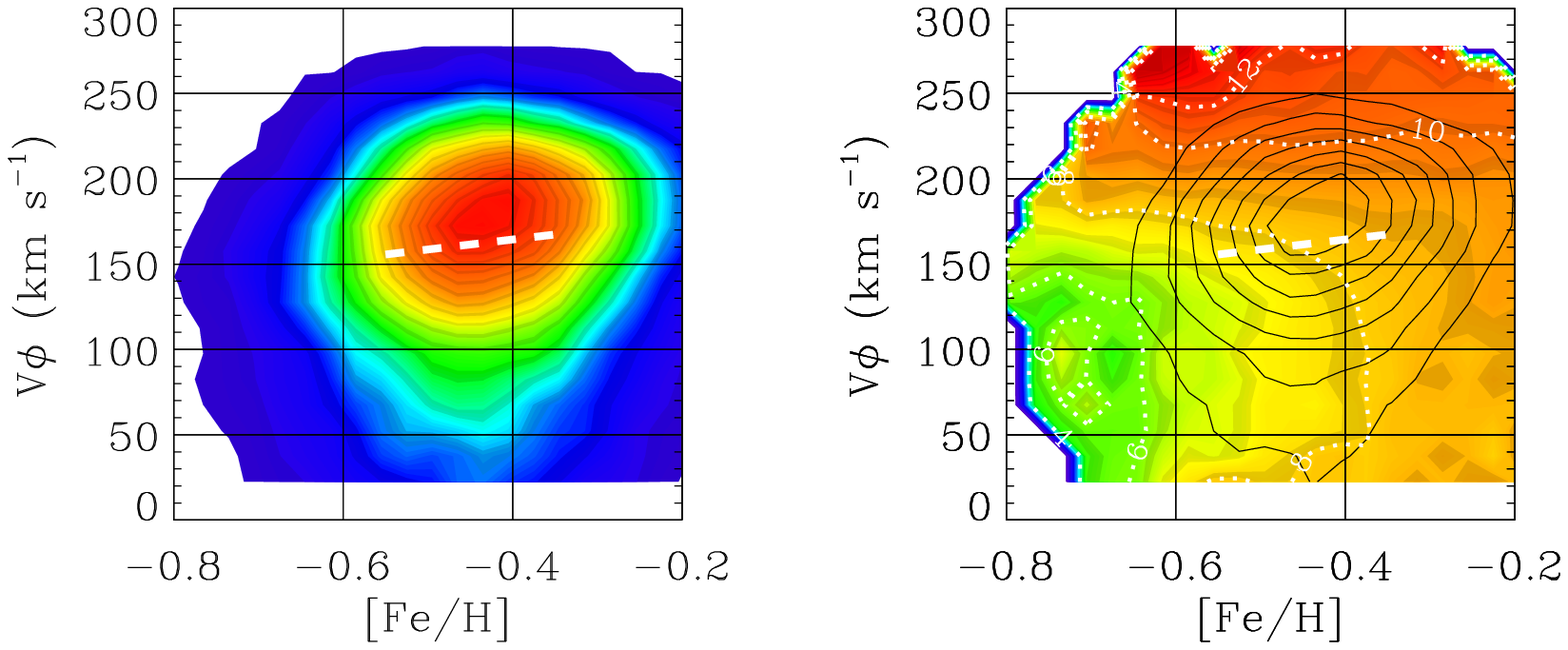}
    \includegraphics[scale=0.9]{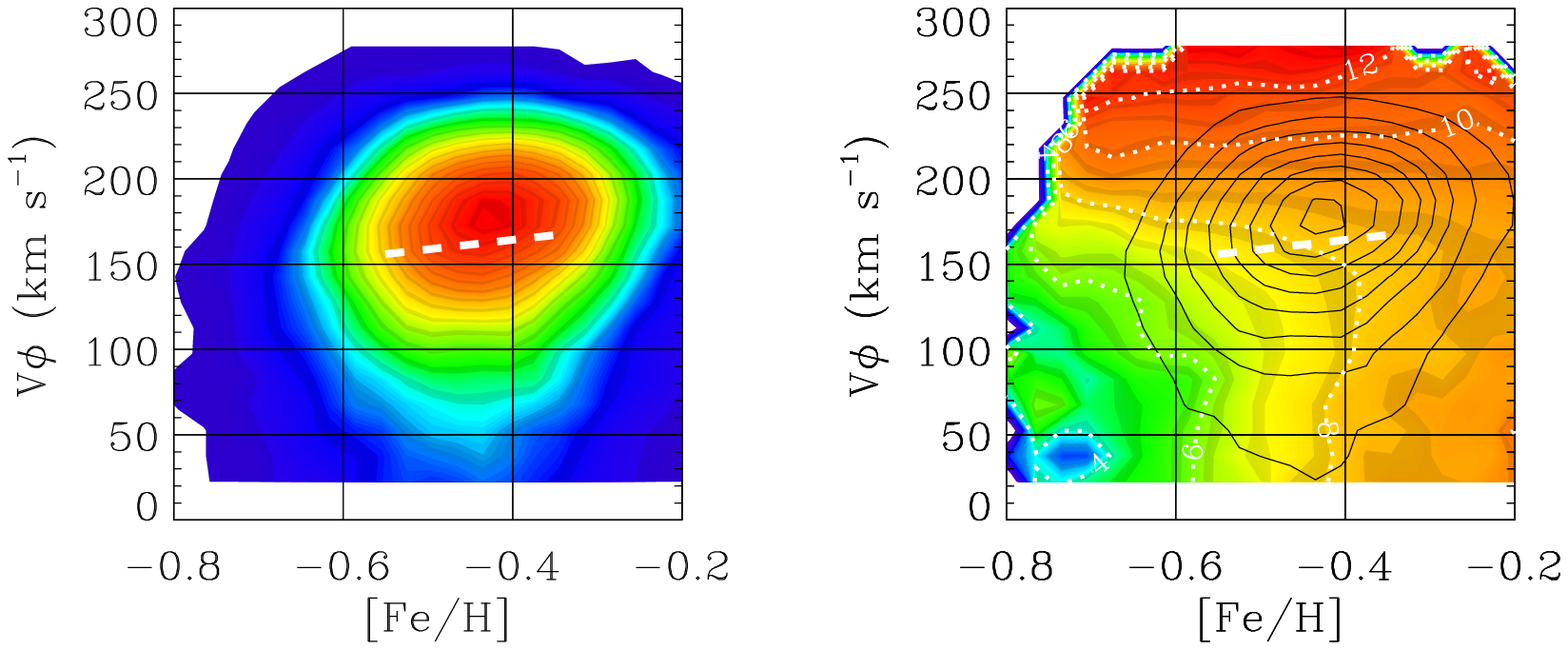}
    \includegraphics[scale=0.9]{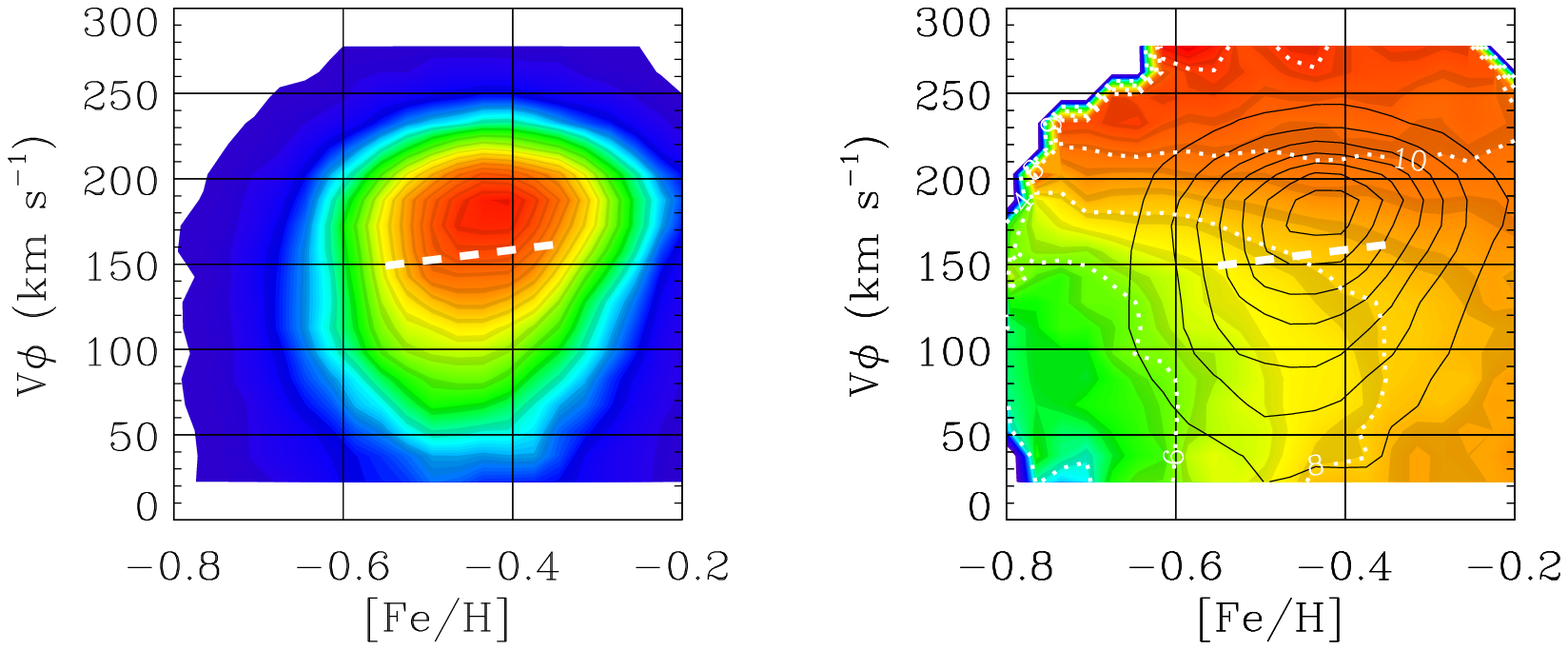} }
\caption{\footnotesize Distribution of the azimuthal velocity and metallicity after an evolution of 1, 3, 5  Gyr. }
\label{fig:correlation_nn}
\end{figure*}

\begin{figure*}
\hspace{2.5truecm} \vbox{ 
    \includegraphics[scale=0.9]{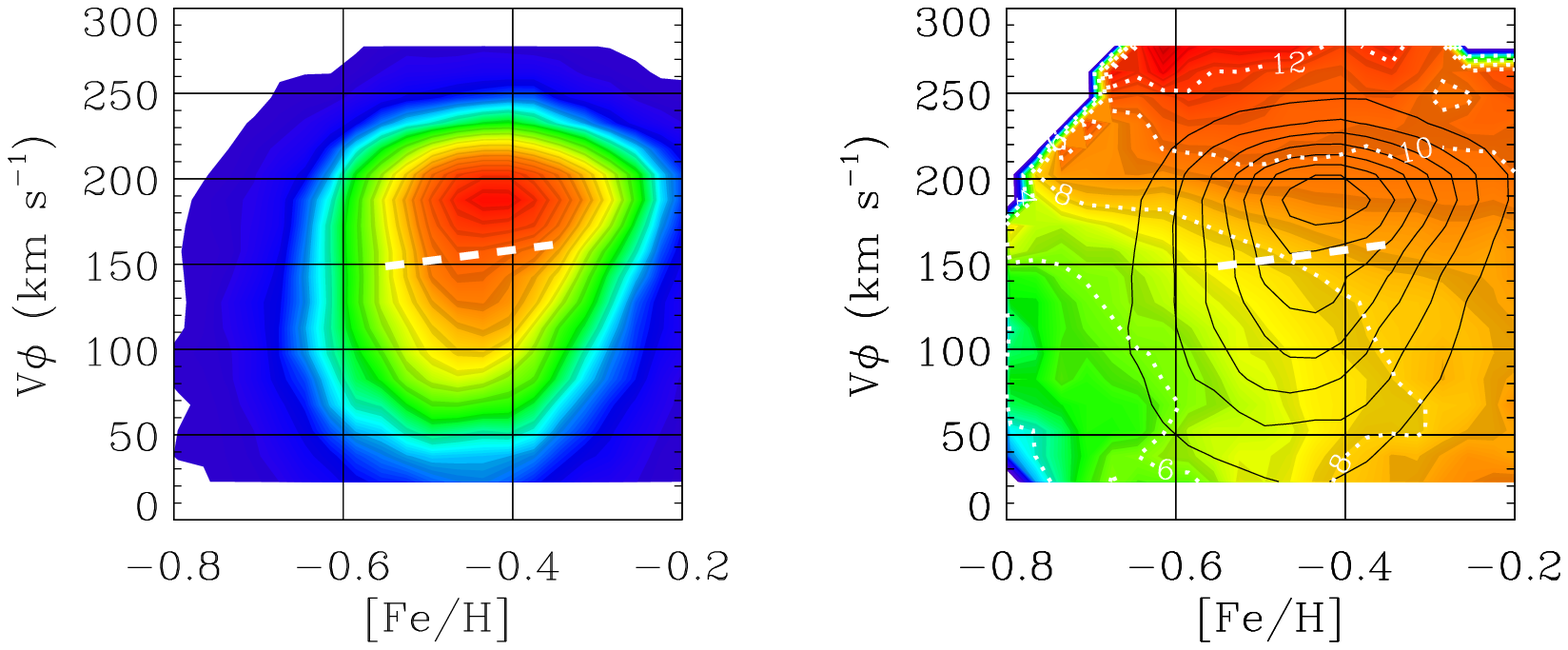}
    \includegraphics[scale=0.9]{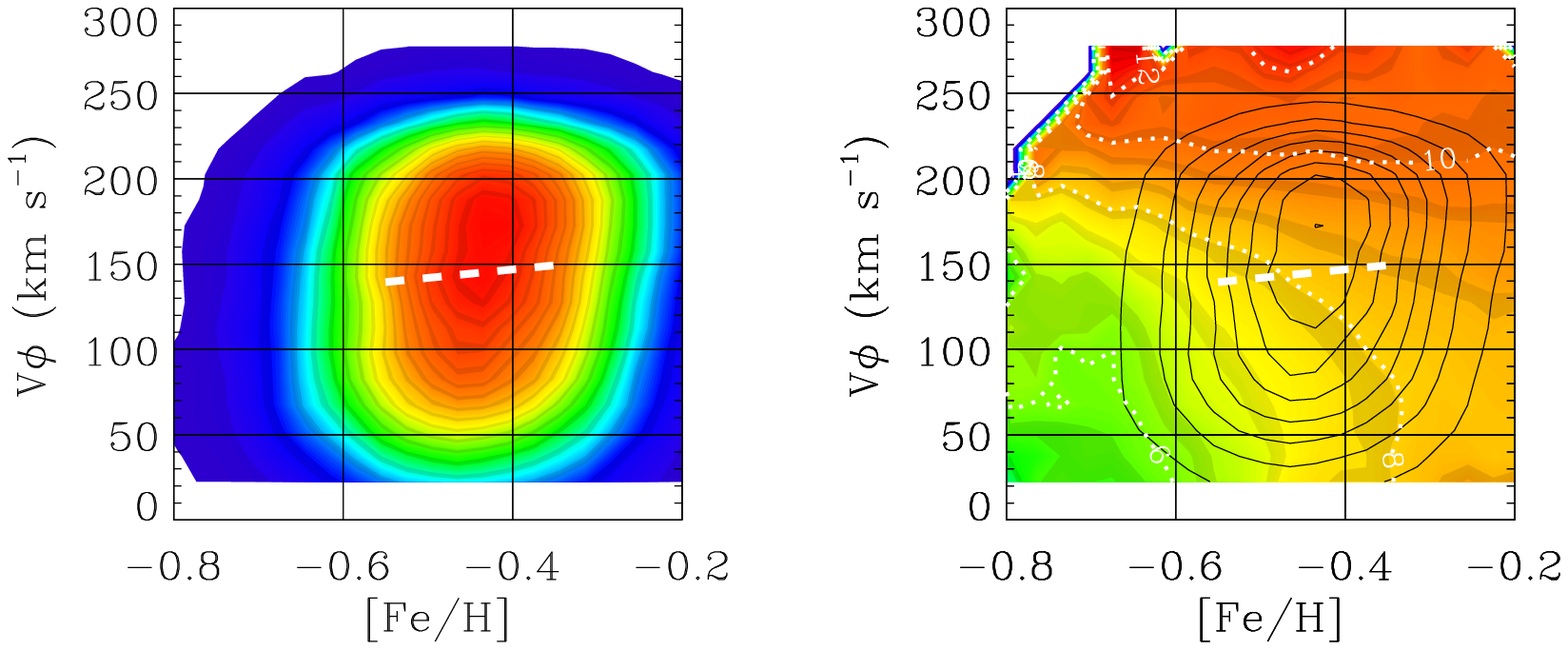}
    \includegraphics[scale=0.9]{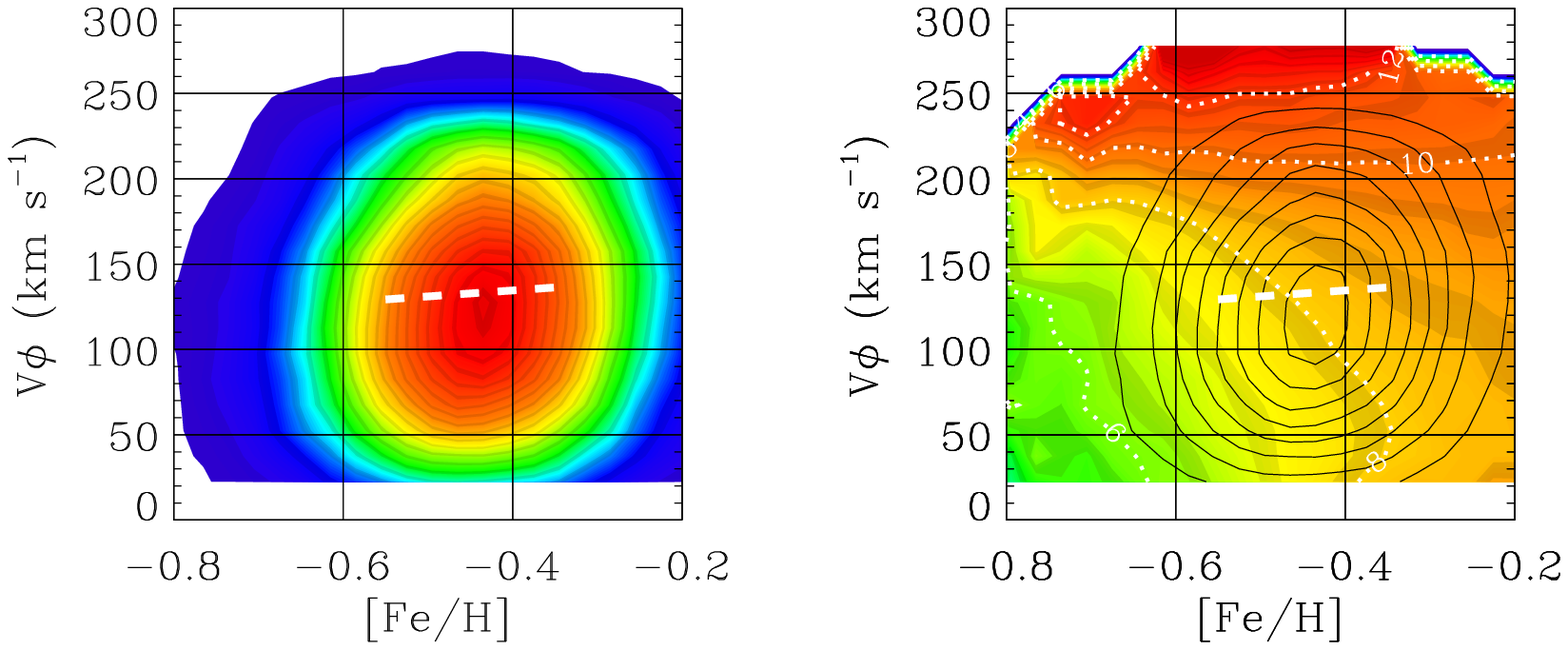} }
\caption{\footnotesize Distribution of the azimuthal velocity and metallicity after an evolution of 6, 7, 8 Gyr.}
\label{fig:correlation_nnn}
\end{figure*}

 The evolution of the rotation-metallicity correlation $\partial  V_\phi / \partial$[Fe/H] in the barred disk model, evaluated for particles in the range  $-0.55<$ [Fe/H] $< -0.35$ and $ V_\phi > 50$ km~s$^{-1}$,  is shown as a dashed thick line
in Figures \ref{fig:correlation_n}, \ref{fig:correlation_nn} and \ref{fig:correlation_nnn}. This choice for the chemical and kinematical ranges allows one to consider the core of the metallicity distribution and exclude the low-velocity tail.  These figures represent  the distribution $V_\phi$ vs. [Fe/H] at different times within the solar annulus   8 kpc
$< R_0<$ 10 kpc and for 1.5 kpc $ <|z|<$ 2.0 kpc.
 The colour-coded zones shown in the right panels of Figures  \ref{fig:correlation_n}, \ref{fig:correlation_nn} and \ref{fig:correlation_nnn} represent the distribution of the average  $\langle R_0\rangle$ original
radii of the particles as a function of the location on the plane ($V_\phi$,[Fe/H]).\\
Table \ref{corr_time} lists the values of the correlation at various times  and the time evolution of the correlation  is shown in  Figure \ref{ctempo}. \\
By definition, at $t=0$ Gyr all  particles belong to the annulus 8 kpc
$< R_0<$ 10 kpc
and no rotation-metallicity correlation can be  present because the [Fe/H]
values are randomly
drawn and assigned to the particles at that time.
Because of stellar motions, at $t=0.1$ Gyr and in the 8-10 kpc annulus, we found many
particles coming from the outer and inner regions, although no
rotation-metallicity correlation is  present as yet.

A mild correlation appears between  $t=0.2$ and $t=0.3$
Gyr, which quickly increases up to $82 \pm 6$ km s$^{-1}$ dex$^{-1}$ at $t=0.4$ Gyr (Figures 2 and 3).
We note that this correlation
is produced by the ''immigration'' of low-rotating and metal-poor
particles from
the inner regions, which populate the bottom-left corner of the right
panels (as indicated by the dotted white iso-contours) and tilt
the iso-density contours  in the left panels. 
As shown in Figure  \ref{fig:correlation_nn}, the rotation-metallicity distribution is stable over several Gyr, until about 6 Gyr. 
As reported in Table \ref{corr_time}, the number of particles is 
also nearly constant
at $\approx$ 30\,000 in the 8-10 kpc annulus and 1.5 kpc $<|z|<2.0$ kpc.

After 6 Gyr, we observe a significant broadening of the $V_\phi$ distribution due
to the arrival of many
particles from the inner regions of the disk that populate the
slowly rotating tail of the velocity
distribution ($V_\phi\la 100$ km s$^{-1}$).  We note, however, that a
smaller and decreasing
rotation-metallicity correlation persists up to $t=8$ Gyr (cfr.
Table \ref{corr_time}).
Notice also that the number of particles increases dramatically at 7 Gyr, reaching $\sim
100\,000$ at $t=8$ Gyr, with a corresponding dramatic change in the azimuthal velocity distribution, as shown in the last two panels of Figure \ref{hist}.\\
This figure helps following the migration history of the particles of interest by showing in  the left panels the histograms of their original radial distributions at the same evolutionary times as those used for Figures  \ref{fig:correlation_n},   \ref{fig:correlation_nn} and \ref{fig:correlation_nnn}. Clearly the peak of the distribution moves with the evolution of the disk outside of the  solar annulus towards smaller initial radii. 
\begin{figure*}
 \centering
  \includegraphics[angle=-90, width=0.9\linewidth]{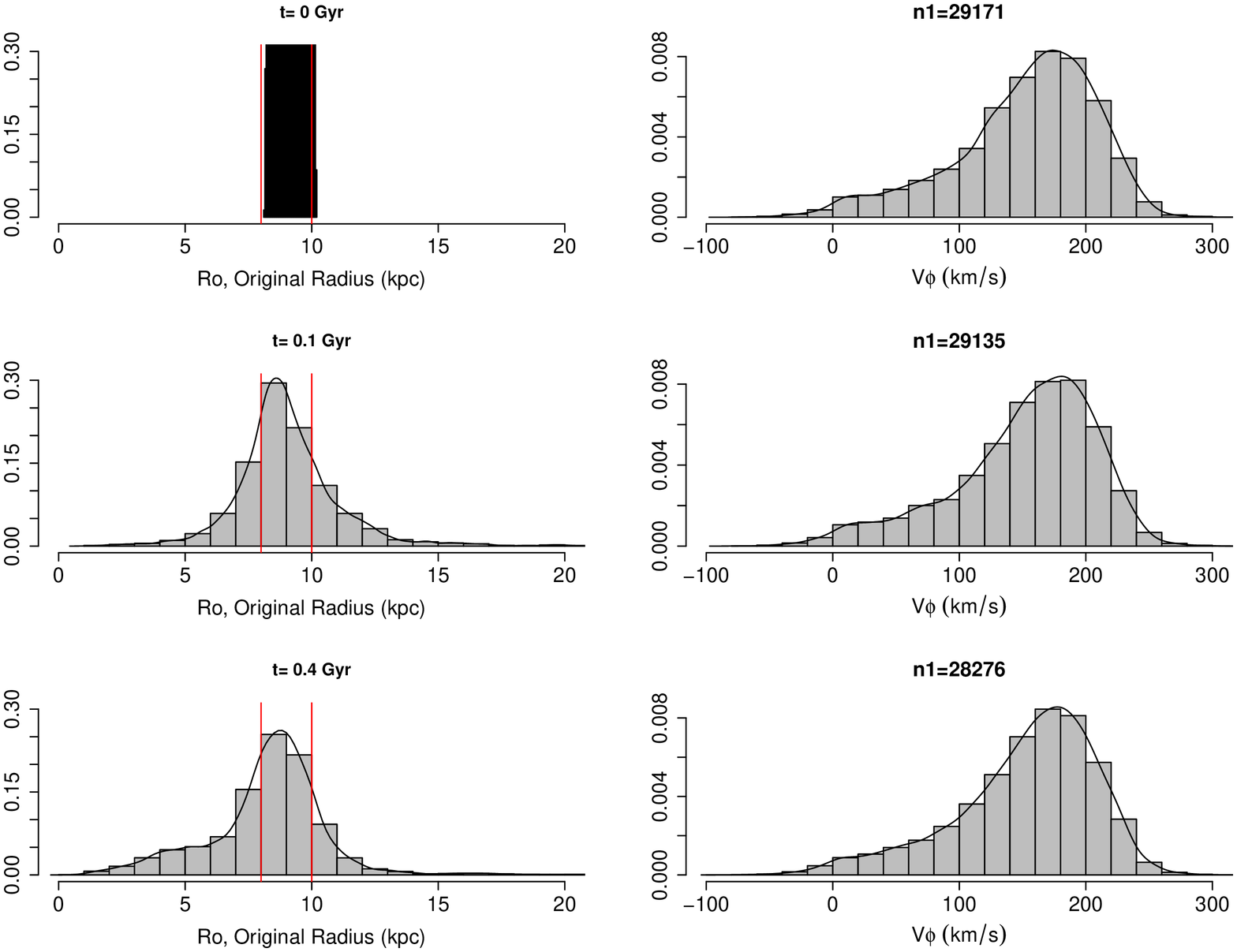}
  \includegraphics[angle=-90, width=0.9\linewidth]{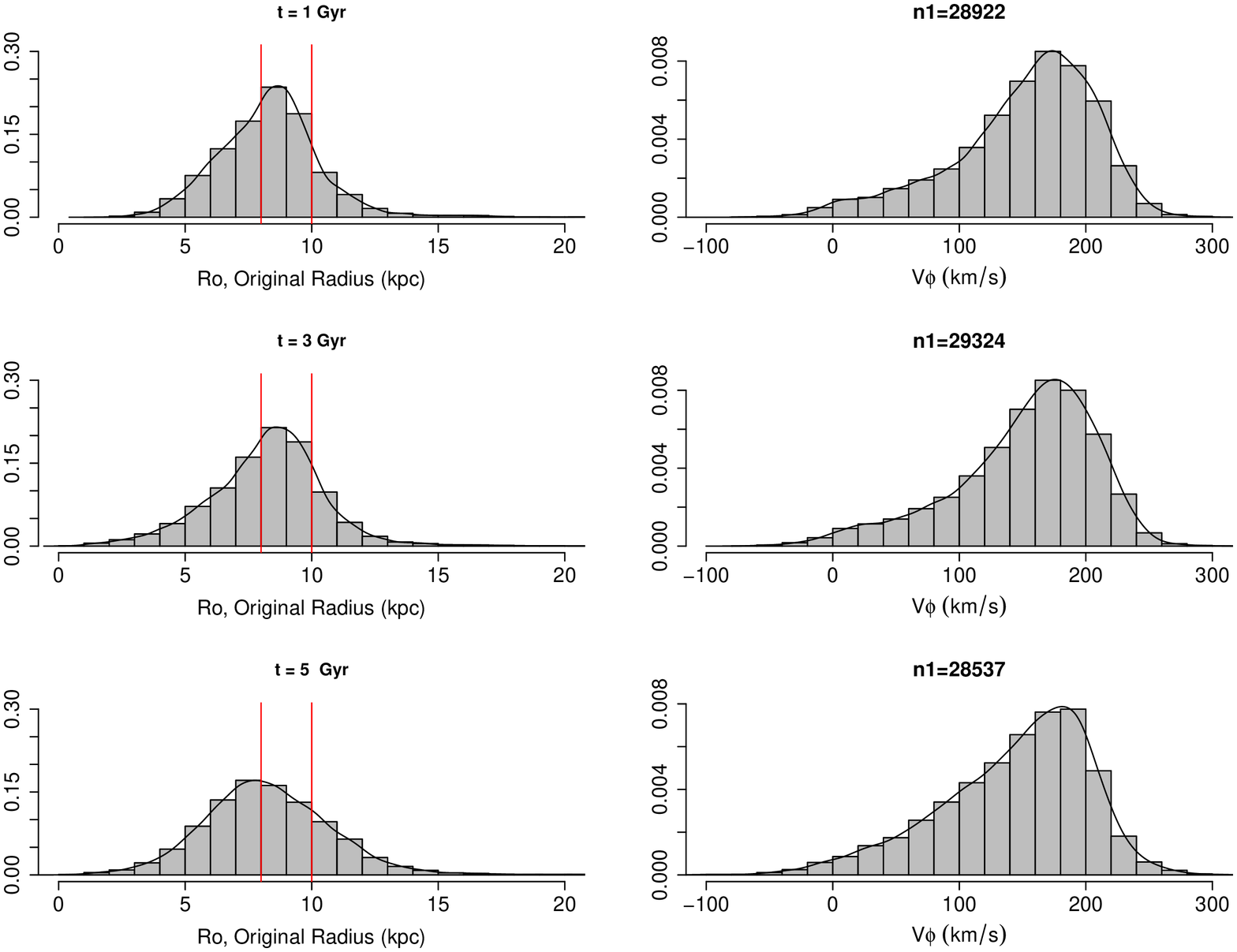}
\end{figure*}
\begin{figure*}
\centering
  \includegraphics[angle=-90, width=0.9\linewidth]{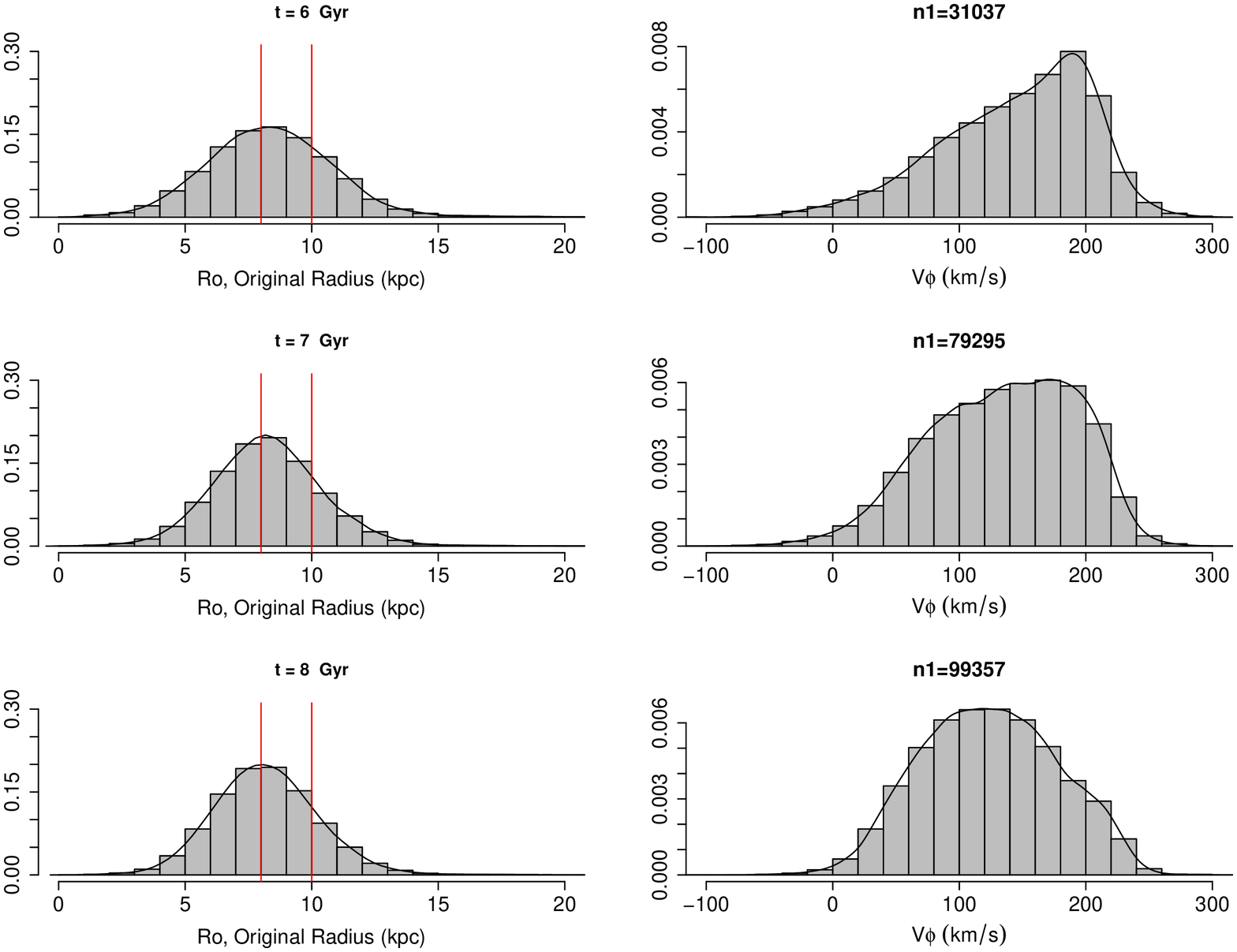}
\caption{ {\it Left panels}: Normalised histograms of the original radii of the subset of particles found within 8 kpc $<$ R $<$ 10 kpc  that have  1.5 kpc $< |z| <$ 2 kpc at different evolutionary times. 
{\it Right panels}: Normalised distributions of the azimuthal velocity at the same times.}
\label{hist}
\end{figure*}

 Finally, it is worth to remark on the intriguing feature of the initial sharp peak in Figure \ref{ctempo}.  We estimated an average timescale of $\approx$ 0.2 Gy for the stars involved to oscillate around their guiding centres, a timescale compatible with the time needed for the formation of this initial spike in the evolution of the rotation-metallicity correlation.  A deeper investigation on the role of the orbital features and on the dynamical orbital evolution will be the subject of a forthcoming study.

\subsection{Barred vs unbarred disks, impact of metalliticy dispersion, and  impact of relaxation}
\begin{figure}
\resizebox{\hsize}{!}{\includegraphics{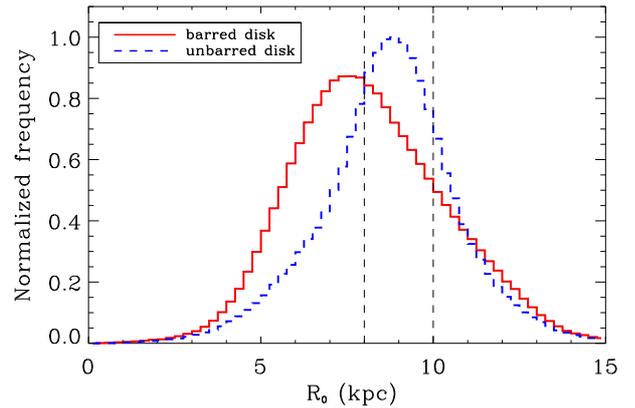}}
\resizebox{\hsize}{!}{\includegraphics{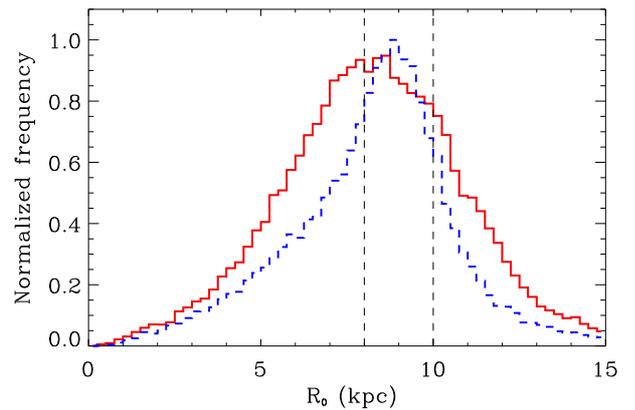}}
\caption{\footnotesize Upper panel: histogram of the original radial distributions of the subset of  particles found within  8 kpc $< R< 10$ kpc after 5 Gyr. Lower panel: histogram of the original radial distributions  of the same particles as in the upper panel, but with  1.5 kpc
  $<|z|< 2.0$ kpc.}
\label{fig:r0distribution}
\end{figure}

\begin{figure*}
\resizebox{\hsize}{!}{\includegraphics[angle=-90]{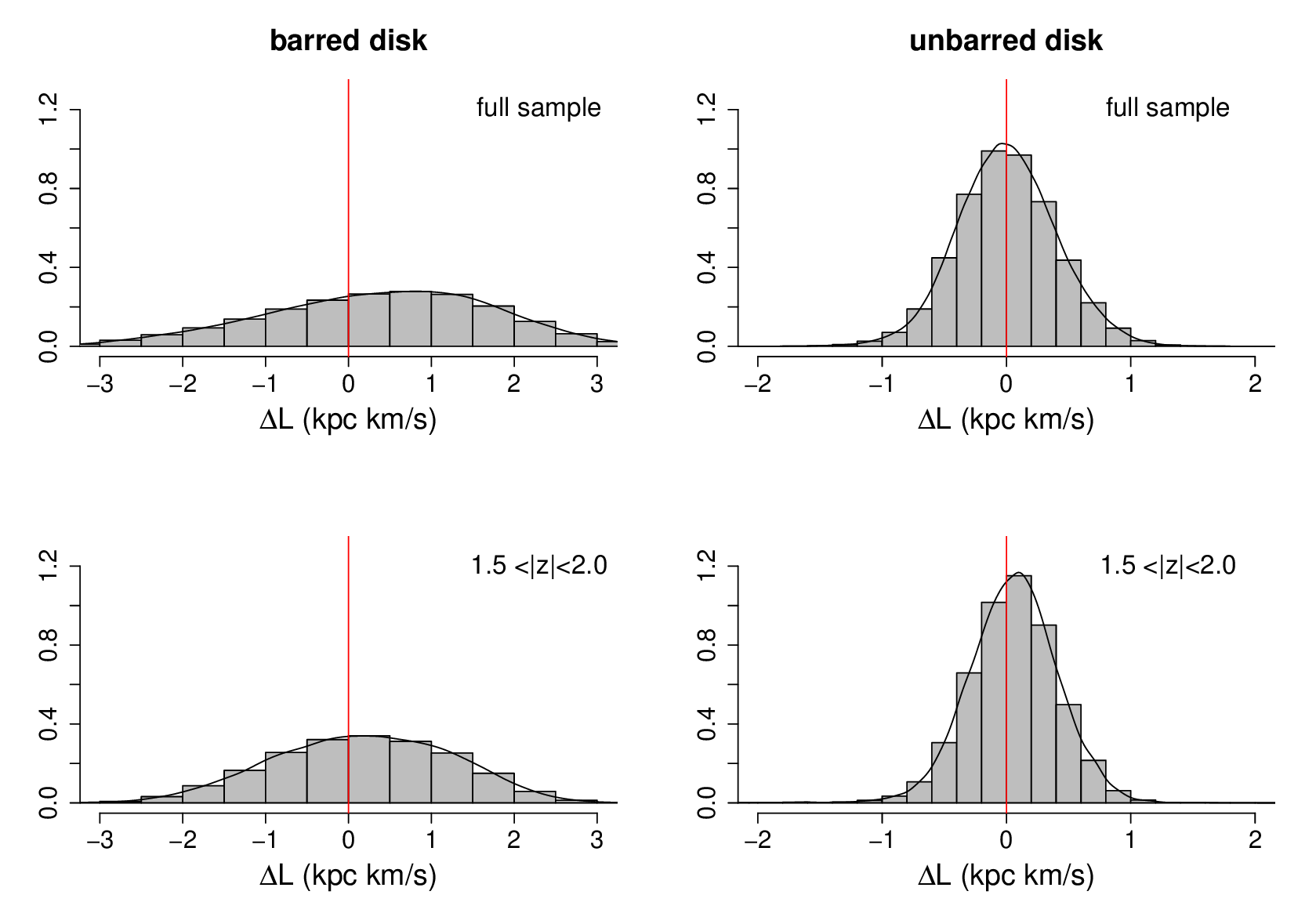}}
\caption{\footnotesize
Normalised distributions of the differences $L_{fin}-L_{ini}$ between initial and final total angular momenta (after 5 Gyr) of the particles inside the solar annulus for the barred and unbarred disk (the curves represent an interpolation of the histogram). }
\label{hist_ang}
\end{figure*}

Figure 
\ref{fig:r0distribution}, upper panel, shows the original radial distribution 
of only those particles that after 5 Gyr are found within the solar annulus (8 kpc$< R< 10$ kpc), whereas the lower panel shows the subset of these particles that have  1.5 kpc
  $<|z|< 2.0$ kpc. 
  Here we can see how  the outward radial migration  
is stronger in the barred disk: there, the bar is driving the
migration  since, according to  \cite{friedli1994} and \cite{minchev2010}, the presence of  nonaxisymmetric features in the disk causes radial migration. On the
other hand, it appears that a diffusion is also produced by the
coarse-grained gravitational field present in our simulations,
as  indicated by the results of our nonbarred disk case.
A fraction of this diffusion could be numerical; however, it
might mimic the diffusion generated by large molecular clouds
in real galaxies (\cite{wielen77},
\cite{jenkins1990}).\\
In the barred-disk case the bulk of
the particles at $R=8$-10 kpc arrives from the inner region of the disk (Figure \ref{fig:r0distribution}, upper panel). Because these particles
 originated in internal regions affected by a stronger vertical potential,
they can extend to higher heights off the plane (Figure \ref{fig:r0distribution} , lower panel).
These results are consistent with  previous studies carried out by
\cite{roskar2008} and \cite{loebman2011}.\\ 
 In Figure \ref{hist_ang} we show the histograms describing the distribution of the differences $L_{fin}-L_{ini}$ between the initial and final, after 5 Gyr, angular momenta of the selected particles, in the solar annulus $8-10$ Kpc, for the  barred and unbarred disks. In the top histograms, we present the angular momenta differences for all  particles in the annulus, whereas    the lower panels show     the same differences for these particles with heights above the plane in the interval 1.5 kpc $<|z|< $2 kpc.
 These plots show how  angular momentum exchanges are  more important for the barred case. The less evident skewness of the histogram for the barred case and  particles at higher $|z|$ reflects the more symmetric distribution shown by the red curve in  the lower panel of Figure \ref{fig:r0distribution}.

To test the effect of  metallicity dispersion on our results, we repeated our numerical experiment by plugging the same metallicity profile, but with different initial metallicity dispersions. As expected,  the rotation--metallicity correlation is strongly dependent on the assumed  initial metallicity dispersion. Indeed, at t=5 Gyr its value becomes 21$\pm$ 8  km s$^{-1}$ dex$^{-1}$  if we run our model with an initial metallicity dispersion of 0.2 dex, while it becomes 157 $\pm $ 6  km s$^{-1}$ dex$^{-1}$ with an initial metallicity dispersion set  to 0.05 dex.\\
 We note that the impact on the rotation-metallicity correlation of a higher initial metallicity dispersion (a reduction in the correlation) appears to be similar to that caused by churning, which might be dominating after 6 Gyr (see Figure \ref{ctempo}). This will be investigated in our forthcoming study.

 Finally, to address the question of the possible role of partial/incomplete dynamical relaxation, we 
  re-assigned the initial metallicities after 2 Gyr of evolution and compared
  the results to the 1-Gyr case. No
  significant difference was found, thus confirming that our results are not
  the spurious effect of relaxation of the adopted initial conditions. 

\subsection{Comparison to the data}

  The results of Figure 2 testify to a positive rotation-metallicity correlation that persists at $\simeq $ 60 km s$^{-1}$ dex$^{-1}$ for several Gyr, to be compared with the 40-50 km s$^{-1}$ dex$^{-1}$ measured by \cite{spagna2010} and \cite{lee2011}.\\
 Although we did not attempt any rigorous  fit to the data, which would not be appropriate given the limitations of our galaxy model, we  emphasise the consistency  of the kinematics produced by our dynamical simulations with what was unveiled by the data.\\
\begin{figure}
\resizebox{\hsize}{!}{\includegraphics[angle=90]{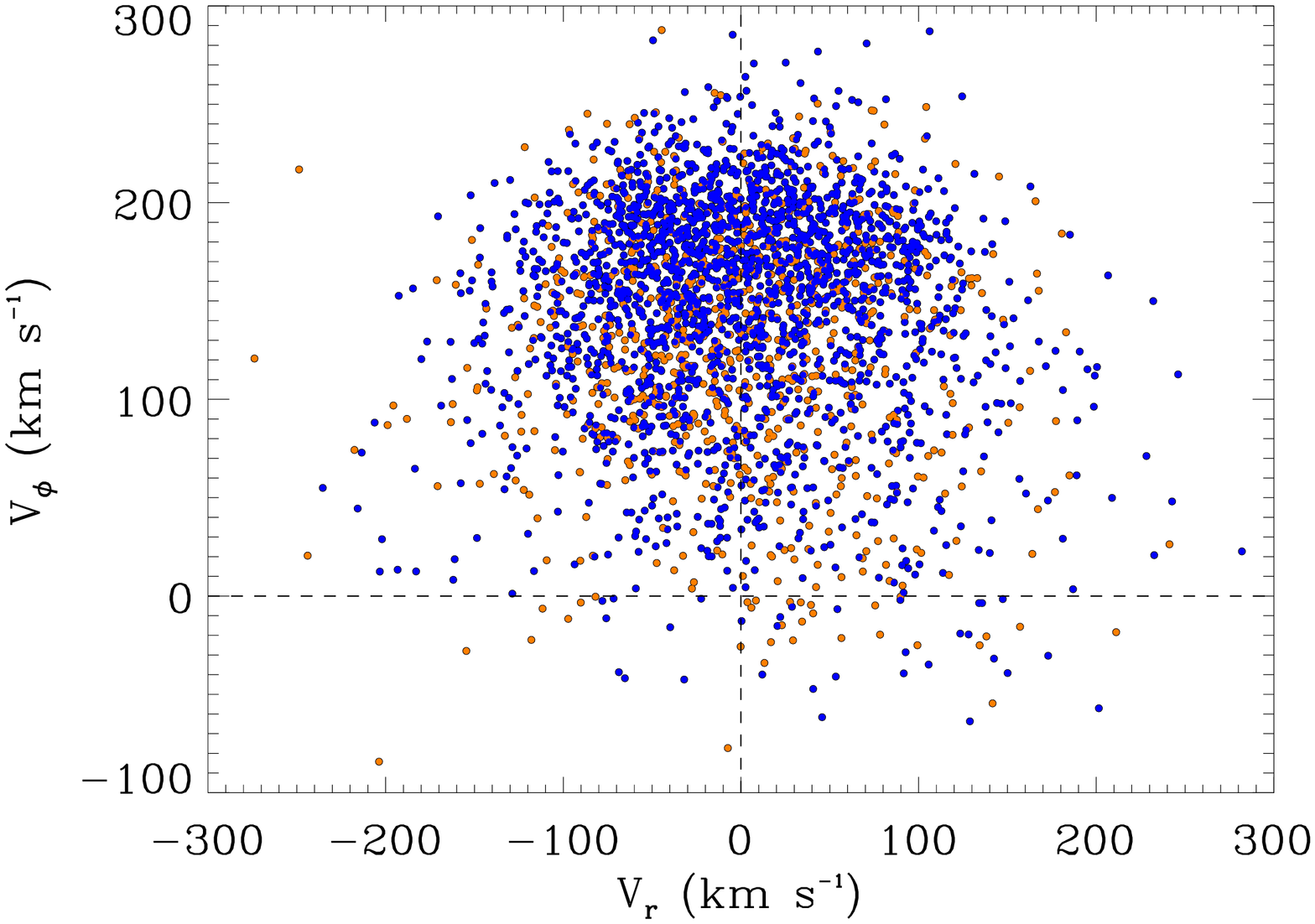}}
\caption{\footnotesize Simulated distribution of the azimuthal velocity vs. radial velocity based on  particles (8 kpc$< R<$ 10 kpc and 1.5 kpc $<|z|<$ 2.0 kpc) from the barred disk model at $t=5$ Gyr. Here, blue and red symbols mark the particles with simulated [Fe/H]$>-0.5$ and [Fe/H]$\le -0.5$, respectively.}
\label{fig:VphiVradAnna}
\resizebox{\hsize}{!}{\includegraphics[angle=90]{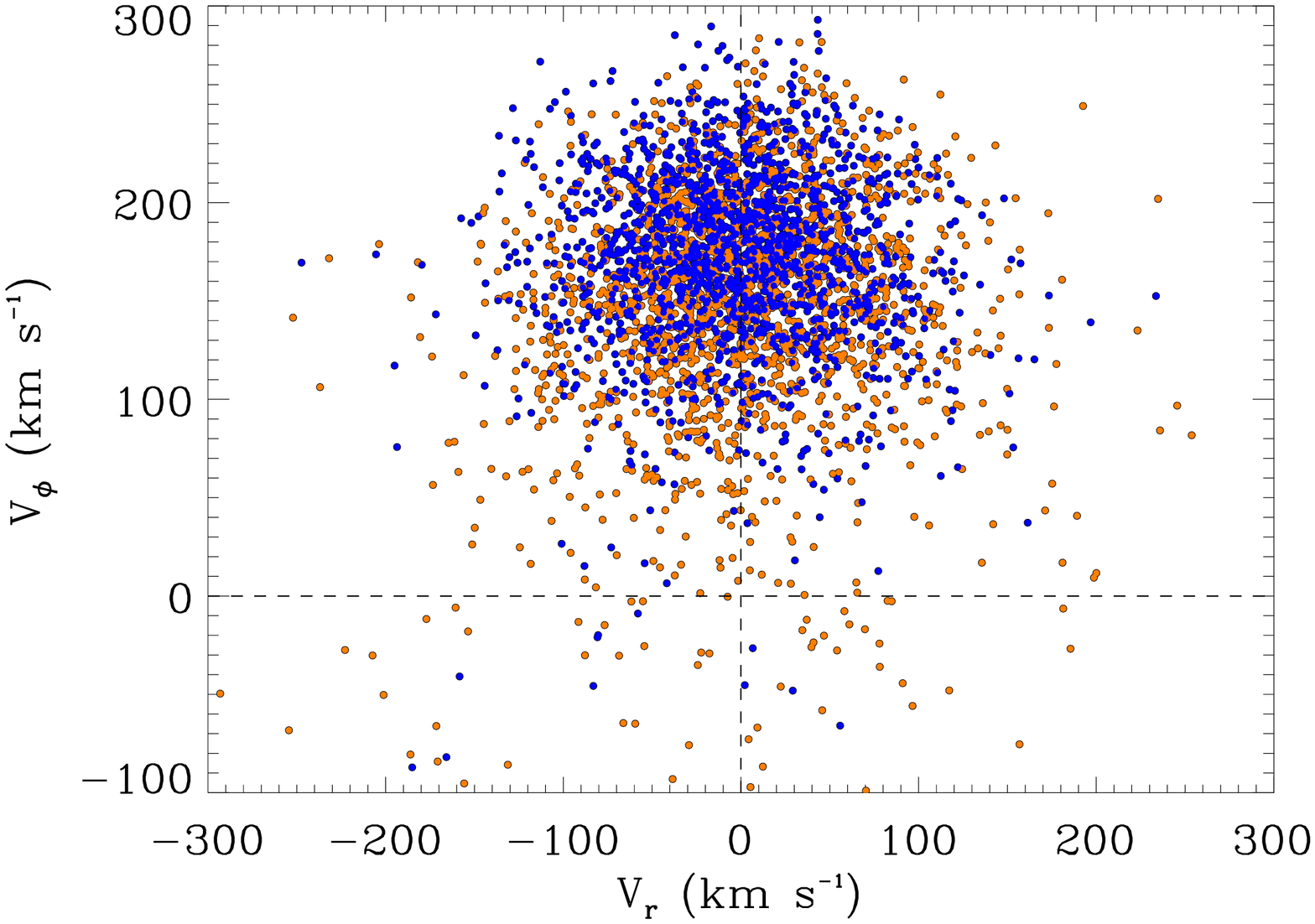}}
\caption{\footnotesize Distribution of the azimuthal velocity vs. radial velocity based on 3507 stars of the SDSS -- GSC-II catalog (1.5 kpc $<|z|<$ 2.0 kpc 
and $-1.0 \le \rm{[Fe/H]}< -0.3$).
Here, blue and red symbols mark the stars with measured [Fe/H]$>-0.6$ and [Fe/H]$\le -0.6$, respectively. }
\label{fig:VphiVradPaola}
\end{figure}
Figures \ref{fig:VphiVradAnna}  and \ref{fig:VphiVradPaola} compare the velocity distribution,  $V_\phi$ vs. $V_R$, of the $N$-body particles in the peripheral 
$z$-layers  to the stellar distribution derived from the SDSS -- GSC-II catalogue studied by \cite{spagna2010}.
The theoretical distribution shows particles from the adopted  barred-disk model evolved for 5 Gyr and selected within the solar annulus at 
1.5 kpc $<|z|<$ 2.0 kpc. 
We note that, in spite of the model limitations, the two
distributions appear to be quite similar,  the moments of the simulated
distribution ($\langle V_\phi \rangle \simeq 142$ km s$^{-1}$, $\sigma_{V\phi}\simeq 59$ km s$^{-1} $,
$\sigma_{V_R}\simeq 70$ km s$^{-1} $) are quite close to the values of the observed distribution 
($\langle V_\phi \rangle \simeq 159$  km s$^{-1}$, $\sigma_{V\phi}\simeq 56$ km s$^{-1} $, $\sigma_{V_R}\simeq
62$  km s$^{-1}$. \\
In addition, we note  an excess of slowly rotating particles ($V_\phi \la $ 100
km s$^{-1}$) with respect to the observations; these are  particles
coming from the innermost regions of the disk, and this  excess seems to
depend on the  bar generated in the simulated disk, which is stronger than
  the MW bar and likely more efficient in  displacing particles.
\section{Discussion and conclusions}
 Nondissipative  mechanisms inside disks due to an irregular
gravitational field  as already theorised by
\cite{wielen77}  are very efficient in  promoting  stellar orbital diffusion. 
 Nonaxisymmetric structures such as spiral arms and bars {promote significant radial migration} (\citep{Sellwood2002} and \citep{minchev2010}).
\cite{schonrich2009a} discussed in detail the {\it churning} and {\it
  blurring} processes that are responsible for the radial mixing. 
Through the comparison between  barred and  unbarred disk models, our simulations confirm  the orbital diffusion in both cases but a more efficient radial mixing in the presence of a bar.
These dynamical processes, which naturally produce  secular migration of the
stars from the disk central regions  towards larger radii and higher vertical distances from the  plane, represent a plausible mechanism
to explain the origin of the MW thick disk.\\
We  recall that our simulations do not include any gaseous component or star formation. We simply investigated the 
evolution of a system formed by collisionless particles representing the 5-10 years old parent population of the thick disk component we observe today.
Through this approach, we dramatically reduced the number of degrees of freedom of the model, 
so that we were able to investigate the details of the dynamical processes 
and their effects on the thick disk chemical properties {\it in situ}, i. e., at higher heights from the plane, where we  assume that no significant contamination due the younger thin disk stars is present.     In the light of our results, the thick disk rotation-metallicity correlation \citep{spagna2010, lee2011} seems to represent an important signature of the 
disk evolution. This feature can indeed be explained by the role of  the
dynamical migration in  our $N$-body simulations,
once we assume an initial radial chemical gradient as that suggested by the
chemical evolution models of \cite{spit01}, which
prescribes a positive slope for the inner early disk, $R\la 10$ kpc, combined with the usual decreasing slope in the outer disk.  The crucial role of a positive inner slope for the primordial chemical distribution to produce a positive rotation-metallicity correlation was presented in \cite{curir},  where a distribution consisting of two simple linear functions with positive slope for inner radii up to 6 kpc and a negative one   outwards was plugged in the early (N-body) disk. \footnote{ This proves that the radial location of the peak of the metallicity distribution adopted in this paper has no  relevance  for the formation of the rotation-metallicity correlation.}\\
\begin{table}
\caption[Table 4]{Evolution of the rotation--metallicity correlation}
 \begin{tabular}{*{4}{c}}
  \hline $t$  & $ n1 $ &  $\partial V_\phi/\partial[Fe/H]$ \\
\hline (Gyr) &    & ( km~s$^{-1}$ dex$^{-1}$)\\
\hline   0.0 &  $ 29171 $  & $3 \pm 6 $\\
  0.1  & 29135 &    $6\pm 6$ \\
  0.25 & 28592 &    $22 \pm 6$\\
   0.35 & 28676 &   $36 \pm 6$\\
   0.4  & 28276 &    $82 \pm 6$\\       
   0.5 & 28926 &   $70 \pm 6$\\
   0.8  & 28928      &  $66 \pm 6$  \\
   1.0  & 28922 &  $58 \pm 6$  \\
   2.0   & 32943 &  $64 \pm 6$ \\
   3.0  & 29324  &  $56 \pm 6$ \\
   4.0 & 30062 &  $68\pm 6$ \\
   5.0 & 28537 &  $62 \pm 6$\\
   6.0 & 31037 &    $ 63 \pm 7 $\\
   7.0 &  79295 &   $ 49 \pm 4$\\
   8.0  &  99357 &   $ 35 \pm 4$\\

\hline
  \end{tabular}
\begin{list}{}{}
\item[Notes.] $t$:  time of evolution. $n1$ :number of particles having 1.5 kpc $<|z|< $2.0 kpc.
\end{list}
\label{corr_time}
\end{table}
The scenario above  is  consistent with the recent results published by
\cite{cresci2010}, who studied a sample of distant galaxies at redshift $z>3$
and found evidence for an {\it inverse} metallicity gradient, possibly
produced by the accretion of primordial gas, which diluted the abundance of
 elements heavier than helium in the centre of the galaxies. Thus, as the
negative correlation,  $-22$ km s$^{-1}$ dex$^{-1}$, of the thin disk
\citep{lee2011} results from the current -- monotonically decreasing -- radial
gradient,  the positive rotation-metallicity correlation of about 40-50 km
s$^{-1}$ dex$^{-1}$, shown by the ancient thick disk population, can be explained by the relic signature of the inverse chemical gradient of the early ISM from which these stars  formed 8-12 Gyr ago.
\begin{acknowledgements}
A.C. and P.R.F. acknowledge the financial support of the Italian Space Agency
through ASI contracts I/037/08/0  and  I/058/10/0  (Gaia Mission - The
Italian Participation to DPAC). A.C. and G.M.  acknowledge the financial support
of INAF through the PRIN 2010 grant n. 1.06.12.02: ``Towards an Italian
Network for Computational Cosmology''. The simulations were carried out at CASPUR, with CPU time assigned through the
Standard HPC grant 2010 : ``Secular processes in disk galaxies and the
formation of the thick disk''.\\
We thank Alvaro Villalobos for his contribution in providing the initial condistions of the numerical simulations.\\
 Thanks are due to the referee Ralph Sch\"{o}nrich for several comments,  which helped us improve the manuscript.
\end{acknowledgements}

\bibliographystyle{aa}

\begin{thebibliography}{}
\bibitem[Abadi et al.~(2003)]{abadi2003} Abadi, M. G., Navarro, J. F., Steinmetz, M., \& Eke, V.R. 2003, ApJ, 597, 21

\bibitem[Abazajian et al.\ (2009)]{abazajian2009} Abazajian, K.N.,  Adelman-McCarthy, J.K.,  Ag\"{u}eros, M.A., et al. 2009, ApJS, 182, 543







\bibitem[Allende Prieto et al.\ (2008)]{allende2008} Allende Prieto, C., Thirupathi, S., Beers, T.  et al. 2008, AJ, 136, 2070.

\bibitem[Arnadottir et al. (2010)]{arnadottir2010} Arnadottir et al.2010, A\&A, 521, 40

\bibitem[Bond et al. (2010)]{bond2010} Bond, N.A., Ivezi\'{c}, \v{Z}, Sesar, B. et al. 2010, ApJ, 716, 1


\bibitem[Bournaud \& Elmegreen (2009)]{bournaud2009} Bournaud, F. \& Elmegreen, B. 2009, ApJL 694, 158 

\bibitem[Brook et al.\ (2005)]{brook2005}  Brook, C.B., Gibson, B.K., Martel, H., \& Kawata, D. 2005, ApJ, 630, 298



\bibitem[Cescutti et al. (2007)]{cesc07} Cescutti, G., Matteucci, F., François, P., Chiappini, C., 2007, A\&A, 462, 943

\bibitem[Chiappini et al. (2001)]{chiap01} Chiappini, C., Matteucci, F.,  Romano, D., 2001, ApJ, 554, 1044 
\bibitem[Chiappini et al. (1997;2001)]{chiap97}Chiappini, C., Matteucci, F., Gratton, R., 1997, ApJ, 477, 765 

\bibitem[Chiappini(2009)]{chiappini2009} Chiappini, C.\ 2009, IAU 
Symposium, 254, 191 


\bibitem [Cresci et al.~(2010)]{cresci2010} Cresci, G., Mannucci, F., Maiolino, R. et al. 2010, Nature, 467, 811



\bibitem[Curir et al. (2012)]{curir} Curir, A., Spagna, A., Lattanzi, M. G., Murante, G., Re Fiorentin, P. 2012 in {\it Assembly the Puzzle of the Milky Way} EPJ  Web of Conferences 19, 10003

\bibitem[Freeman (1987)]{Freeman 1987} Freeman, K.~C.\ 1987, \araa, 25, 603 

\bibitem[Freeman, \& Bland-Hawthorn (2002)]{freeman2002} Freeman, K. \& Bland-Hawthorn, J. 2002 \araa, 40, 487

\bibitem[Friedli et al. (1994)]{friedli1994} Friedli, D., Benz, W., Kennicutt, R. 1994, ApJ 430, L105







\bibitem[Gratton et al. (2003)]{gratton2003} Gratton, R.G., Carretta, E., Desidera, S. et al. 2003 A\&A, 406, 131
  

\bibitem[Haywood (2008)]{haywood2008}  Haywood, M. 2008, MNRAS, 388, 1175

\bibitem[Hernquist (1993)]{hernquist1993} Hernquist, L.\ 1993, \apjs, 86, 389 

\bibitem[Ivezi\'{c} et al. (2008)]{ivezic2008} Ivezi\'{c}, \v{Z}., Sesar, B.,
 Juri\'{c}, M., et al. 2008, ApJ, 684, 287


\bibitem[Jenkins \& Binney~(1990)]{jenkins1990}
Jenkins, A. \& Binney J. 1990, MNRAS, 245, 305

\bibitem[Kordopatis et al.~(2011)]{kordopatis2011} Kordopatis, G., Recio-Blanco,
  A.,  de Laverny, P., et al. 2011, arXiv:1110.5221







\bibitem[Lasker et al.\ (2008)]{lasker2008} Lasker, B.M., Lattanzi, M.G., McLean, B.J., et al. 2008, \aj, 136, 735



\bibitem[Lee et al.~(2011)]{lee2011} Lee, Y.~S., et al.\ 2011, \apj, 738, 187
  
\bibitem[Loebman et al.~(2011)]{loebman2011} Loebman, S., Ro\v{s}kar, Debattista, V.P. et al. 2011, ApJ, 737, 8
\bibitem[Luck \& Lambert (2011)]{luck2011}  Luck, R. E., Lambert, D. L., 2011, Apj, 142, 136L


\bibitem[Martel et al. (2011)]{martel2011} Martel, H., Richard, S., Brook, C.B., Kawata, D., Gibson, B. K., \& Sanchez-Blazquez, P. (2011) in IAU Symp 277, in press

\bibitem[Matteucci \& Francois (1989)]{matt89}Matteucci, F., Francois, P., 1989, MNRAS, 239, 885 



\bibitem[Minchev \& Famey~(2010)]{minchev2010} Minchev, I. \& Famaey, B. 2010, ApJ, 722, 112

\bibitem[Morrison et al.~(2009)]{morrison2009} Morrison, H. L., Helmi, A., Sun, J., et al. 2009, ApJ, 694, 130


\bibitem[Navarro et al., 1997]{NFW97} Navarro, J.~F., Frenk, C.~S. and White, S.~D.~M. 1997, ApJ, 190, 493


\bibitem[Press et al.~(1986)]{Press1986} Press, W.~H., Flannery, B.~P., \& Teukolsky, S.~A., Vetterling, W.,T.\ 1986, {\it Numerical Recipes}  Cambridge: University Press, 1986  

\bibitem[Quinn et al. (1993)]{quinn1993} Quinn, P.J., Hernquist, L., Fullagar, D.P. 1993, ApJ, 403, 74

\bibitem[Qu et al.(2011)]{Qu2011} Qu, Y., Di Matteo, P., Lehnert, M.~D., \& van Driel, W. 2011, \aap, 530, A10 

\bibitem[Re Fiorentin ~(2012)]{refiorentin2012} Re Fiorentin, P. in: {\it The Chemical Evolution of the Milky Way }, Sexten Center for Astrophysics (SCfA). Sesto Pusteria, Bolzano, Italy, January 23 - 27, 2012


\bibitem[Robin et al. ~(1996)]{robin1996} Robin, A. C., Haywood, M., Creze, M., Ojha, D.K., Bienayme, O. 1996,  A\&A 305, 125 

\bibitem[Ro\v{s}kar et al.~(2008)]{roskar2008} Ro\v{s}kar, R., Debattista, V.P., Stinson, G.S., et al. 2008, ApJ, 675, L65



\bibitem[Sch\"{o}nrich \& Binney~(2009a)]{schonrich2009a} 
Sch\"{o}nrich, R., \& Binney J. 2009, MNRAS, 396, 203

\bibitem[Sch\"{o}nrich \& Binney~(2009b)]{schonrich2009b} 
Sch\"{o}nrich, R., \& Binney J. 2009, MNRAS, 399, 1145

\bibitem[Sellwood \& Binney~(2002)]{Sellwood2002} Sellwood, J.~A., \& Binney, J.~J. 2002, \mnras, 336, 785 




\bibitem[Spagna et al.~(2010)]{spagna2010} Spagna, A., Lattanzi, M.G., Re Fiorentin, P. \& Smart, R.L. 2010, A\&A, 510, L4

 \bibitem[Spitoni \& Matteucci~(2011)]{spit01}Spitoni, E., Matteucci, F., 2011, A\&A,     531, 72 
  
\bibitem[Springel, V.~(2005)]{sprin2005} Springel, V., 2005, MNRAS, 364, 1105


\bibitem[Villalobos \& Helmi~(2008)]{villalobos2008}  Villalobos, \'{A}. \&
  Helmi, A. 2008, MNRAS, 391, 1806


\bibitem[Wielen~(1977)]{wielen77} Wielen, R. 1977, aap, 60, 263 


\bibitem[Yanny et al.\ (2009)]{yanny2009} Yanny, B., Rockosi, C., Newberg, H.J., et al. 2009, AJ, 137, 4377



\end{thebibliography}

\end{document}